\documentclass[runningheads]{llncs}
\usepackage[T1]{fontenc}
\usepackage[utf8]{inputenc}
\usepackage{etoolbox}

\usepackage{etoolbox}

\makeatletter
\patchcmd{\thebibliography}
  {\list}
  {\footnotesize\list}
  {}{}
\makeatother
\usepackage{graphicx}
\usepackage{amsmath,amssymb}
\allowdisplaybreaks

\usepackage{booktabs,multirow,tabularx,array,makecell,threeparttable,adjustbox}
\usepackage{siunitx}
\usepackage[shortlabels]{enumitem}

\usepackage{algorithm}
\usepackage{algorithmicx}
\usepackage{algpseudocode}

\usepackage[hidelinks]{hyperref}

\begin{document}
\title{TeleHunt: A Framework and Tool for Efficient Cybercriminal Community Discovery on Telegram}
\titlerunning{Efficient Cybercriminal Community Discovery on Telegram}
\author{\vspace{-1.5em}}
\institute{\vspace{-1.5em}}
%
\author{
Roy Ricaldi \and
Victor Asanache \and
Luca Allodi
}

\authorrunning{Ricaldi et al.}

\institute{
Eindhoven University of Technology, Eindhoven, The Netherlands\\
\email{\{r.j.ricaldi.saavedra, v.asanache, l.allodi\}@tue.nl}
}

\maketitle              
\begin{abstract}
This paper presents TeleHunt, a framework and tool for evaluating 
the effectiveness of different strategies to discover cybercriminal 
communities on Telegram. TeleHunt employs a set of reference-driven 
snowballing strategies, integrating message-level classification, 
contextual filtering, and market-segment labeling. Using open- and 
dark-web seeds, we systematically evaluate how seed source, pointer 
type, and exploration strategy influence discovery outcomes in three 
dimensions: efficiency, accessibility, and rediscovery. Our work 
provides (i) a modular cybercrime content discovery pipeline, (ii) 
the first systematic comparison of Telegram discovery strategies with 
an empirical characterization of market-segment accessibility, and 
(iii) a labeled dataset of over 172 million messages from 6,022 
Telegram communities.

\keywords{Cybercrime \and Telegram \and Threat Intelligence \and 
Underground Markets \and Snowball Sampling \and Content Classification}
\end{abstract}

\section{Introduction}

Telegram has evolved into a major coordination hub for cybercriminal 
activity, offering anonymity, low barriers to entry, and infrastructure 
well suited for distributing illicit offerings\cite{Allodi2024dmitry,ricaldi_marjanov_hutchings_allodi_2025,darkgram}. Unlike traditional 
dark-web forums, Telegram communities form and disappear rapidly, rely 
heavily on message-based advertisements, and employ varying access 
controls ranging from public to private or vetted communities.

While tools for cybercrime detection on Telegram exist, practitioners 
lack an understanding of \emph{how different exploration strategies 
perform} and \emph{what their outcomes reveal about the Telegram 
cybercrime ecosystem}. Existing work on Telegram cybercrime primarily 
focuses on content classification and ecosystem characterization rather 
than systematically evaluating discovery strategies or seed-dependent 
measurement bias, leaving practitioners without evidence-based guidance 
on how to structure efficient collection efforts 
\cite{darkgram,Hughes2024science,Campobasso2022threatcrawl}. What is 
missing is a systematic framework to evaluate cybercriminal community 
discovery strategies and enable targeted CTI hunting on Telegram.

To address this gap, we develop \emph{TeleHunt}, a modular 
language-model-driven pipeline that automates the discovery of 
cybercriminal communities through iterative snowballing of messages 
with \emph{pointers} to other communities. We evaluate discovery using 
three metrics: \emph{efficiency} (how productively new communities are 
identified), \emph{accessibility} (how open or restricted the ecosystem 
is), and \emph{rediscovery} (the extent to which exploration revisits 
known communities). These metrics allow us to assess not only which 
strategies perform best, but also what discovery patterns reveal about 
the structure of Telegram's cybercrime economy.

Our contributions are as follows:
\begin{itemize}
    \item \textbf{TeleHunt pipeline.} We present a configurable, 
    LLM-assisted framework integrating scraping, pointer extraction, 
    contextual filtering, iterative expansion, and market-segment 
    labeling for scalable cybercriminal community discovery. The 
    pipeline builds on the taxonomy and classifiers from prior work 
    \cite{ricaldi_marjanov_hutchings_allodi_2025}; our contribution 
    is the discovery framework, its systematic evaluation, and the 
    resulting dataset.

    \item \textbf{Systematic evaluation of discovery strategies.} We 
    provide a reproducible comparison of seed sources, pointer types, 
    and contextual filtering, identifying effective CTI hunting 
    strategies and offering the first empirical assessment of Telegram 
    cybercrime ecosystem accessibility.

    \item \textbf{Labeled dataset and release.} We collect 6{,}022 
    communities (3{,}471 cybercriminal), comprising 172{,}385{,}463 
    messages from 2{,}392{,}741 users, annotated with market segment, 
    community type, size, and temporal attributes; the dataset can be 
    shared for research purposes.
\end{itemize}

Our study is guided by four research questions: \textbf{RQ0 
(TeleHunt):} How can we design a framework to evaluate cybercriminal 
community discovery strategies? \textbf{RQ1 (Efficiency):} Which 
techniques most efficiently uncover cybercriminal communities? 
\textbf{RQ2 (Accessibility):} What do discovery outcomes reveal about 
market-segment accessibility? \textbf{RQ3 (Rediscovery):} To what 
extent does discovery repeatedly surface already-seen communities?

\section{Related Work and Background}
\label{sec:related_work_background}

Prior work has measured underground forums, markets, and Telegram 
communities to understand the scale, structure, and economics of 
cybercrime ecosystems 
\cite{Marjanov2025sok,article,Marjanov2024breaking,ricaldi2026topicalshiftsdarkweb}. 
Crawling frameworks such as THREAT/crawl allow systematic data 
collection from these environments \cite{Campobasso2022threatcrawl}, 
supporting studies that analyze attacker preferences 
\cite{Campobasso2023know,Hughes2024science,Hughes2023digital,Anderson2012measuring}, 
community relevance within the broader threat landscape 
\cite{Campobasso2023company,CABREROHOLGUERAS2021102489,Pastrana2018characterizing,10.1145/2464464.2464524}, 
cross-platform migration \cite{Allodi2024dmitry}, and the 
infrastructure foundations of criminal ecosystems 
\cite{10.1093/bjc/azab026}.

On Telegram, identifying illicit communities is challenging due to 
high message volume, informal language, and platform-specific slang 
\cite{hughes}. The DarkGram study \cite{darkgram} analyzed 339 
cybercrime Telegram communities using seeded collection and LLM-based 
classification, demonstrating large-scale distribution of illicit 
content across six market segments. TeleHunt differs from DarkGram in 
that it systematically compares discovery strategies rather than 
characterizing a fixed collection, and evaluates how seed source, 
pointer type, and traversal configuration affect what is found and how 
efficiently. Similarly, Garkava et al.\ \cite{article} examine stolen 
data markets on Telegram but focus on crime script analysis rather 
than discovery methodology. Large Language Models have improved 
automated classification of underground content \cite{darkbert,hanxiang}, 
and fine-tuned models achieve high accuracy in market-segment 
categorization \cite{darkgram}. However, existing work primarily 
focuses on content classification and ecosystem characterization 
rather than on evaluating discovery strategies.

Measurement of hidden populations typically relies on seed selection 
and snowball sampling \cite{1521d86f-33df-3ece-8080-69edbcaba312}. 
Seed diversification across dark- and open-web sources mitigates 
thematic bias \cite{darkgram}, while iterative expansion supports 
network exploration. Yet no prior study systematically evaluates how 
seed origin, pointer type, contextual filtering, or traversal 
configuration affect discovery efficiency or ecosystem coverage on 
Telegram.

Telegram communities function as groups or channels and vary in 
accessibility: public (searchable and handle-based), private 
(invite-link based), or vetted (administrator approval required). 
Prior work identifies six market segments---Cyberattacks, Digital 
Infrastructure, Digital Piracy, Fraud Tools, Personal Data, and 
Tutorials---supporting structured ecosystem analysis 
\cite{ricaldi_marjanov_hutchings_allodi_2025}. What remains unclear 
is how different discovery strategies expose these segments and access 
models, whether discovery saturates through rediscovery, and how 
configuration choices shape observable ecosystem structure.
\section{Methodology}
\label{sec:methodology}

\begin{figure}[t]
    \centering
    \includegraphics[width=0.98\linewidth]{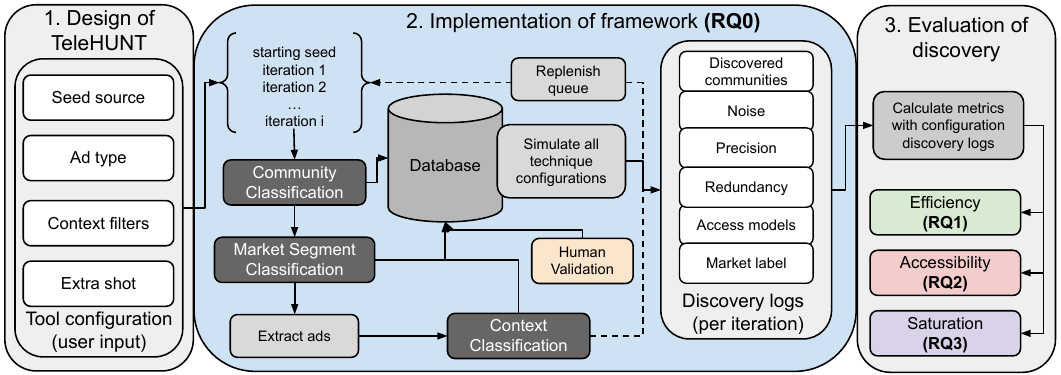}
    \caption{Methodology overview.}
    \label{fig:meth}
\end{figure}

\autoref{fig:meth} summarizes the three stages of our methodology: 
(i) framework design, (ii) implementation, and (iii) evaluation. 
Supplementary materials including the handle reliability analysis, 
database schema, and an anonymized data sample are available in our 
online repository at {\url{https://github.com/royricaldi/telehunt}}.

\subsection{TeleHunt Design}

TeleHunt automates cybercriminal community discovery from an initial 
seed and user-defined configuration. It scrapes communities via the 
Telegram API, classifies messages using fine-tuned language models, 
extracts advertised community pointers, and iteratively expands 
through snowballing.

Configurations combine: (i) seed source, (ii) pointer type, (iii) 
contextual filtering, and (iv) expansion parameters, summarized in 
Table~\ref{tab:config-legend}. All configurations were executed for 
three iterations to enable systematic comparison.\footnote{We 
restricted expansion to three iterations to limit temporal drift in 
community composition, as deeper traversal increases exposure to 
time-sensitive pointer decay and platform churn, leading to wasted 
resources and low marginal discovery. Sensitivity of results to this 
choice is left for future work; we note that dark-web configurations 
already show rising per-iteration redundancy by iteration~3, 
suggesting saturation may emerge under further expansion.} Human 
validation of community classification is performed per seed origin 
on a sample of 100 communities to confirm output validity 
(Appendix~\ref{app:humanvalidation}). To identify which techniques 
materially affect discovery of \emph{valuable} communities, we 
estimate a Beta regression on the proportion of valuable discoveries 
per pointers processed.\footnote{Beta regression was selected because 
the dependent variables (e.g., yield, precision, noise) are continuous 
proportions bounded in the open interval $(0,1)$, allowing flexible 
mean--variance structure.} Only statistically significant components 
are retained for detailed evaluation.

We evaluate configurations along three dimensions: \emph{Efficiency} 
(discovery rate, yield, precision, noise), \emph{Accessibility} 
(public/private/vetted distribution and exposed market segments), and 
\emph{Rediscovery} (rate of repeated community encounters).

\begin{table}[t]
\centering
\small
\caption{Configuration components.}
\setlength{\tabcolsep}{3pt}
\renewcommand{\arraystretch}{1.0}
\begin{tabular}{ll}
\toprule
\textbf{Seed}      & D: Dark web \quad O: Open web \\
\textbf{Pointers}  & H: Handle \quad L: Link \quad F: Forward \\
\textbf{Context}   & C: Criminal only \quad B: Both contexts \\
\textbf{Expansion} & X: Extra shot \quad N: No extra shot \\
\bottomrule
\end{tabular}
\label{tab:config-legend}
\end{table}

\subsection{Implementation}

\begin{figure}[t]
    \centering
    \includegraphics[width=0.98\linewidth]{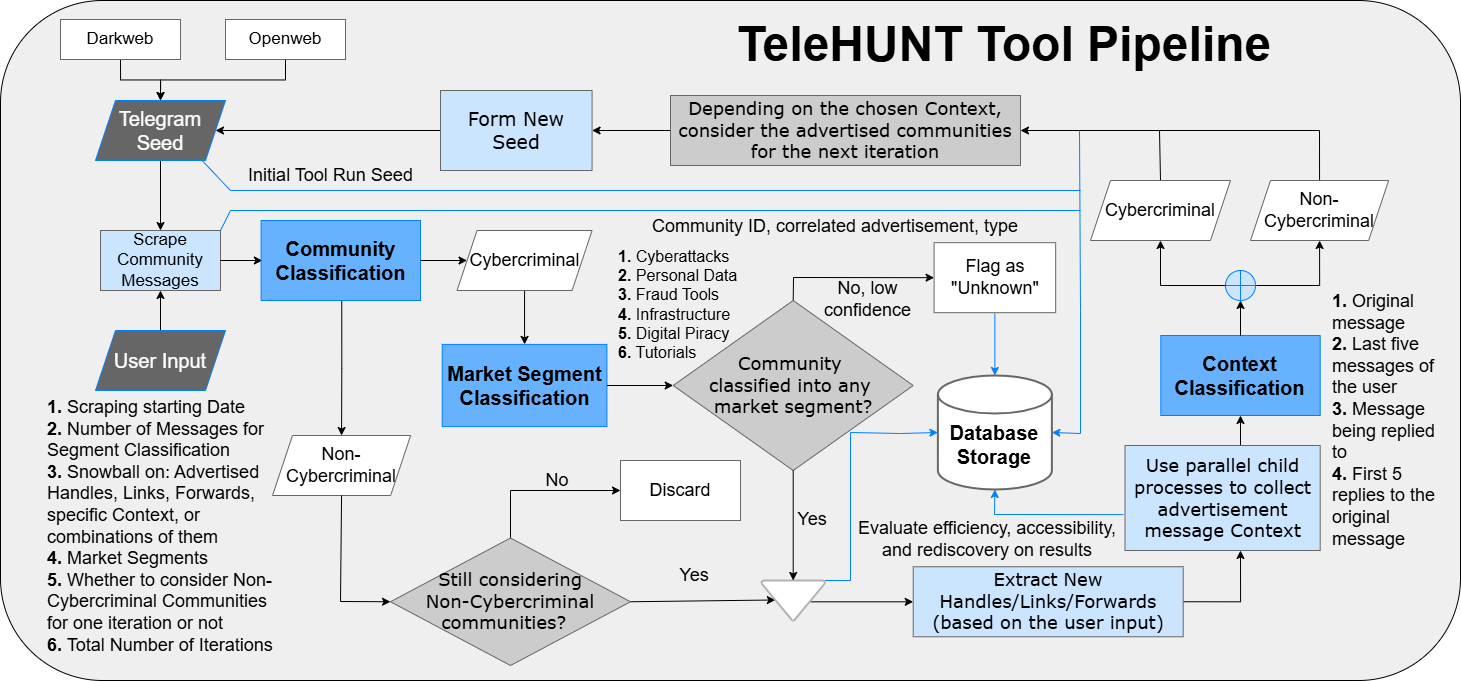}
    \caption{TeleHunt pipeline.}
    \label{fig:framework}
\end{figure}

TeleHunt is implemented as a modular pipeline, detailed in 
Figure~\ref{fig:framework}. Classifier training details, model 
performance metrics, and human validation results are provided in 
Appendix~\ref{appendix:classification}. We use two seed sets of 
cybercriminal communities:\footnote{Originally 50 communities per 
origin; some were taken down before the tool was run.} 41 
communities sourced from threads across various dark-web forums, and 
45 communities from the open web indexed via TGstat.

\subsubsection{Message and Community Classification}

For each iteration, we scrape all text messages within a 30-day 
window, together with metadata (author, timestamp, replies, forwards). 
Communities with fewer than 30 messages in this window are treated as 
dormant and excluded from further processing.

Message-level classification uses two fine-tuned RoBERTa-Large models: 
a binary cybercriminal classifier and a six-class market segment 
classifier (Cyberattacks, Digital Infrastructure, Digital Piracy, 
Fraud Tools, Personal Data, 
Tutorials~\cite{ricaldi_marjanov_hutchings_allodi_2025}). Both achieve 
validation F1-scores above 0.95 (Appendix~\ref{appendix:classification}). 
Each message receives a binary label and, if cybercriminal, a segment 
label. Community-level labels are derived by aggregation: communities 
with $\geq$70\% cybercriminal messages are marked \emph{valuable}, 
and the dominant market segment is assigned by majority vote. Access 
type (public, private, vetted) is inferred at scraping time based on 
join mechanism and API response. Groups and channels are treated 
uniformly as \emph{communities}.

\subsubsection{Pointer Extraction and Contextual Filtering}

From classified communities, we extract three pointer types: handles 
(\texttt{@username}), invite links (\texttt{t.me/+hash}), and forwards 
(origin channel metadata). Handles originating from groups are 
excluded due to high noise: manual verification of handles from 17 
channel-based and 9 group-based dark-web seed communities showed that 
while resolution rates were similar (75\% channel-derived, 76\% 
group-derived), a large fraction of group-derived handles resolved to 
individual user accounts rather than communities, which cannot serve 
as expansion nodes. Full details of this analysis are in our online 
repository.

For each pointer, we construct a context window consisting of the 
pointer message, up to five prior messages by the same author, the 
replied-to message, and up to five replies (maximum eleven messages). 
The window is classified message-by-message; pointers are retained if 
at least 50\% of context messages are labeled cybercriminal. Verified 
pointers form the seed of the next iteration. Rediscovered communities 
are recorded but not reprocessed.

\textbf{Thresholds} were chosen based on empirical distributional 
properties of the data and operational considerations. The 70\% 
valuable-community cutoff ensures that communities are predominantly 
illicit while tolerating minor non-criminal content. The 50\% 
contextual filter balances precision and recall in pointer expansion. 
The 30-day scraping window captures recent activity while limiting 
temporal drift. The dormancy cutoff ($<$30 messages) filters inactive 
communities; manual inspection confirmed that communities below this 
threshold were overwhelmingly abandoned rather than new or emerging. 
Sensitivity of findings to these threshold choices is left for future 
work.

\subsubsection{Execution and Logging}

All runs are logged in an SQLite database storing seeds, scraped 
communities, classifications, pointers, and rediscovery events; the 
full schema is in our online repository. Each iteration extracts 
pointers, applies contextual filtering, removes duplicates, scrapes 
newly discovered communities, and repeats for up to three expansion 
iterations. This logging design enables full reproducibility of all 
yield, precision, noise, and rediscovery metrics reported in 
Section~\ref{sec:results}.

\subsubsection{Human Validation}

Before calculating discovery metrics, we audit classification on a 
stratified sample of 100 communities per dataset (dark web and open 
web), reviewing up to 30 messages per community and comparing human 
annotations with automated labels using exact agreement and Cohen's 
$\kappa$. Results show substantial agreement and are documented in 
Appendix~\ref{app:humanvalidation}.

\subsection{Evaluating TeleHunt}
\label{sec:evalmeth}

Using the discovery logs generated during exploration, we evaluate 
TeleHunt along three dimensions. All metrics are summarized in 
Table~\ref{tab:metrics}.

\subsubsection{Efficiency}
\label{sec:efficiency}

Efficiency measures how productively exploration converts pointers 
into new and \emph{valuable} communities, defined as active 
communities with $\geq$70\% cybercriminal messages. Visited 
communities may yield pointers leading to new, valuable, or failed 
outcomes.\footnote{Pointers may fail if the community was deleted, 
banned, or expired, or if the pointer was malformed.} We evaluate: 
(i)~\textit{Discovery rate} (new communities per pointers processed), 
(ii)~\textit{Yield} (valuable communities per pointers processed), 
(iii)~\textit{Precision} (valuable communities per new communities), 
and (iv)~\textit{Noise} (failed pointers per pointers processed). 
Metrics are computed per configuration and per iteration to capture 
both aggregate performance and exploration dynamics.

\subsubsection{Accessibility}
\label{sec:accessibility}

Accessibility reflects the structural openness of discovered 
communities. Each unique community is labeled public, private, or 
vetted at scraping time. We compute the distribution of access types 
per configuration and examine which market segments become reachable 
under each strategy. A higher proportion of public or private 
communities indicates broad discoverability, while a larger vetted 
share suggests restricted regions that may contain higher CTI value 
but limit further expansion.

\subsubsection{Rediscovery}
\label{sec:rediscovery}

Rediscovery measures whether exploration continues to uncover novel 
communities or begins to saturate. We compute per-iteration 
redundancy — the fraction of community-leading pointers pointing to 
already-seen communities — and the cumulative rediscovery rate across 
iterations. Increasing redundancy indicates diminishing marginal 
returns and structural saturation of reachable communities under a 
given configuration.

\subsubsection{Selecting Configurations}
\label{sec:selecting}

Given the large configuration space, we first identify which 
techniques significantly affect discovery of valuable communities 
before conducting detailed evaluation. We model the proportion of 
valuable discoveries ($V/a$) using Beta regression, with predictors 
including pointer type (handles, links, forwards), seed source (dark 
vs.\ open web), contextual filtering (criminal-only vs.\ both), and 
extra shot (Table~\ref{tab:config-legend}). Techniques are retained 
for detailed evaluation only if their regression coefficients are 
statistically significant ($\alpha = 0.05$) and their central 
estimates do not overlap in effect size, ensuring each retained 
component makes a distinct contribution to the outcome.

\begin{table}[t]
\centering
\small
\caption{Metrics and computation.}
\label{tab:metrics}
\setlength{\tabcolsep}{3pt}
\renewcommand{\arraystretch}{1.0}
\begin{tabular}{p{1.8cm} p{6.4cm}}
\toprule
\textbf{Efficiency} &
Iteration $i$: pointers processed $a_i$, failed $f_i$, new $n_i$, 
rediscovered $r_i$, valuable $V_i$. \\
& Discovery rate: $D_i=\frac{n_{i+1}}{a_i}$. \quad
  Yield: $Y=\frac{V}{a}$, $Y_i=\frac{V_{i+1}}{a_i}$. \\
& Precision: $P=\frac{V}{n}$, $P_i=\frac{V_{i+1}}{n_i}$. \quad
  Noise: $O=\frac{f}{a}$, $O_i=\frac{f_i}{a_i}$. \\
& High $D_i$, $Y_i$, $P_i$ indicate productive expansion; 
  high $O_i$ reflects invalid pointers. \\
\midrule
\textbf{Accessibility} &
Fractions of discovered communities: public $p$, private $q$, 
vetted $v$, with $A=\{p,q,v\}$. Higher $p$ or $q$ suggests 
accessible regions; higher $v$ indicates restricted spaces 
limiting expansion. \\
\midrule
\textbf{Rediscovery} &
Cumulative new: $N_i=\sum_{j\le i}n_j$; rediscovered: 
$R_i=\sum_{j\le i}r_j$. Per-iteration redundancy: 
$\rho_i=\frac{r_i}{n_i+r_i}$. Cumulative: 
$\rho=\frac{R_i}{N_i+R_i}$. Rising $\rho$ signals saturation. \\
\bottomrule
\end{tabular}
\end{table}

\subsection{Ethical Considerations}

The research was approved by the ethics committee of the Department of Math and Computer Science of
Eindhoven University of Technology, under ERB approval no. ERB2021MCS1. Data was collected from publicly accessible communities or communities 
advertised on public forums, stored on encrypted institutional servers, 
and analyzed in aggregate; no individual is de-anonymized. Individual 
informed consent was not feasible at this scale; under the British 
Society of Criminology's Ethics Statement~\cite{britishethics}, it is 
not required for research on publicly available online data.

\section{Results}
\label{sec:results}

We apply TeleHunt to two separate scraping runs, using dark-web (41 
communities) and open-web (45 communities) seeds, each expanded for 
up to three iterations across all configurations. After two weeks, 
this yielded 6,022 unique Telegram communities, of which 3,471 were 
classified as cybercriminal. Using logged discovery traces, we 
reconstruct outcomes for every configuration and rely on Beta 
regression to identify techniques that materially affect 
\textit{yield}.\footnote{Reconstruction assumes independent 
configuration effects and deterministic pointer resolution within a 
run. This is a simplifying assumption for a live snowballing process; 
path dependence and API variability may affect attribution, and 
results should be interpreted accordingly.} We report underperforming 
techniques as well, to contextualize differences across 
\textit{efficiency}, \textit{accessibility}, and \textit{rediscovery}.

\subsection{Significance of Techniques}
\label{sec:technique_significance}

Table~\ref{tab:regressions} reports both regression models. 
\textbf{Pointer type dominates yield}: techniques including links 
produce large, significant improvements. L ($\beta=1.34$) and HL 
($\beta=1.25$) show the strongest effects (both $p<2\times10^{-16}$); 
mixed formats LF ($0.71$) and HLF ($0.62$) also significantly increase 
yield. Forwards and handles alone do not differ from the baseline 
($p>0.48$). The presence of links, not the number of pointer types, 
is the primary driver of discovery success. Non-pointer factors have 
modest effects: open-web seed ($\beta=0.20$, $p=0.001$) and criminal 
context ($\beta=0.11$, $p=0.063$) provide secondary gains; extra shot 
has no detectable effect. We therefore fix criminal context, exclude 
extra shot, and vary seed and pointer type. For subsequent analysis 
we compare top configurations (L, HL), include HLF to evaluate full 
ad usage, and retain HF as baseline.

\textbf{Noise regression}: links reduce the odds of an invalid pointer 
by $\approx$90\% ($\exp(-2.24)\approx0.11$). Each passing week 
increases noise by $\approx$50\% ($\exp(0.38)\approx1.46$), 
confirming pointer freshness as a key operational factor. Pointers 
shared in criminal contexts age better: each additional week makes 
them $\approx$14\% less likely to fail than equally old non-criminal 
pointers, suggesting criminally embedded referrals lead to longer-standing 
communities.\footnote{This interpretation is plausible but not uniquely 
implied by the model; alternative explanations include selection effects 
in which communities are advertised in criminal contexts.}

\begin{table}[t]
\centering
\small
\caption{Beta regression for yield (left, baseline: HF) and logistic 
regression for noise (right).}
\label{tab:regressions}
\setlength{\tabcolsep}{3pt}
\renewcommand{\arraystretch}{0.95}
\begin{tabular}{lrr@{\hspace{10pt}}lrr}
\toprule
\multicolumn{3}{c}{\textbf{Yield (Beta reg.)}} & 
\multicolumn{3}{c}{\textbf{Noise (Logistic reg.)}} \\
\textbf{Predictor} & \textbf{Est.} & \textbf{SE} & 
\textbf{Predictor} & \textbf{Est.} & \textbf{SE} \\
\midrule
Intercept   & $-2.07^{***}$ & 0.11 & Intercept        & $1.79^{***}$  & 0.24 \\
HLF         & $0.62^{***}$  & 0.13 & Handle           & $-0.07$       & 0.16 \\
F           & $0.09$        & 0.13 & Link             & $-2.24^{***}$ & 0.08 \\
H           & $0.02$        & 0.13 & Criminal         & $-0.04$       & 0.24 \\
HL          & $1.25^{***}$  & 0.12 & Age (weeks)      & $0.38^{***}$  & 0.07 \\
L           & $1.34^{***}$  & 0.12 & Criminal$\times$Age & $-0.15^{***}$ & 0.08 \\
LF          & $0.71^{***}$  & 0.12 & & & \\
Open        & $0.20^{***}$  & 0.06 & McFadden $R^2$   & $0.18$ & -- \\
Criminal    & $0.11^{*}$    & 0.06 & & & \\
ExtraShot   & $-0.01$       & 0.06 & & & \\
\midrule
$\phi$             & $123.21^{***}$ & 23.66 & & & \\
$\log\mathcal{L}$  & $104.2$        & --    & & & \\
pseudo-$R^2$       & $0.81$         & --    & & & \\
\bottomrule
\multicolumn{6}{l}{\footnotesize $^{*}p{<}0.1$; $^{**}p{<}0.05$; 
$^{***}p{<}0.01$}
\end{tabular}
\end{table}

\subsection{Discovery Efficiency (RQ1)}
\label{sec:results_efficiency}

\begin{figure}[t]
    \centering
    \begin{minipage}{0.49\columnwidth}
        \centering
        \includegraphics[width=\linewidth]{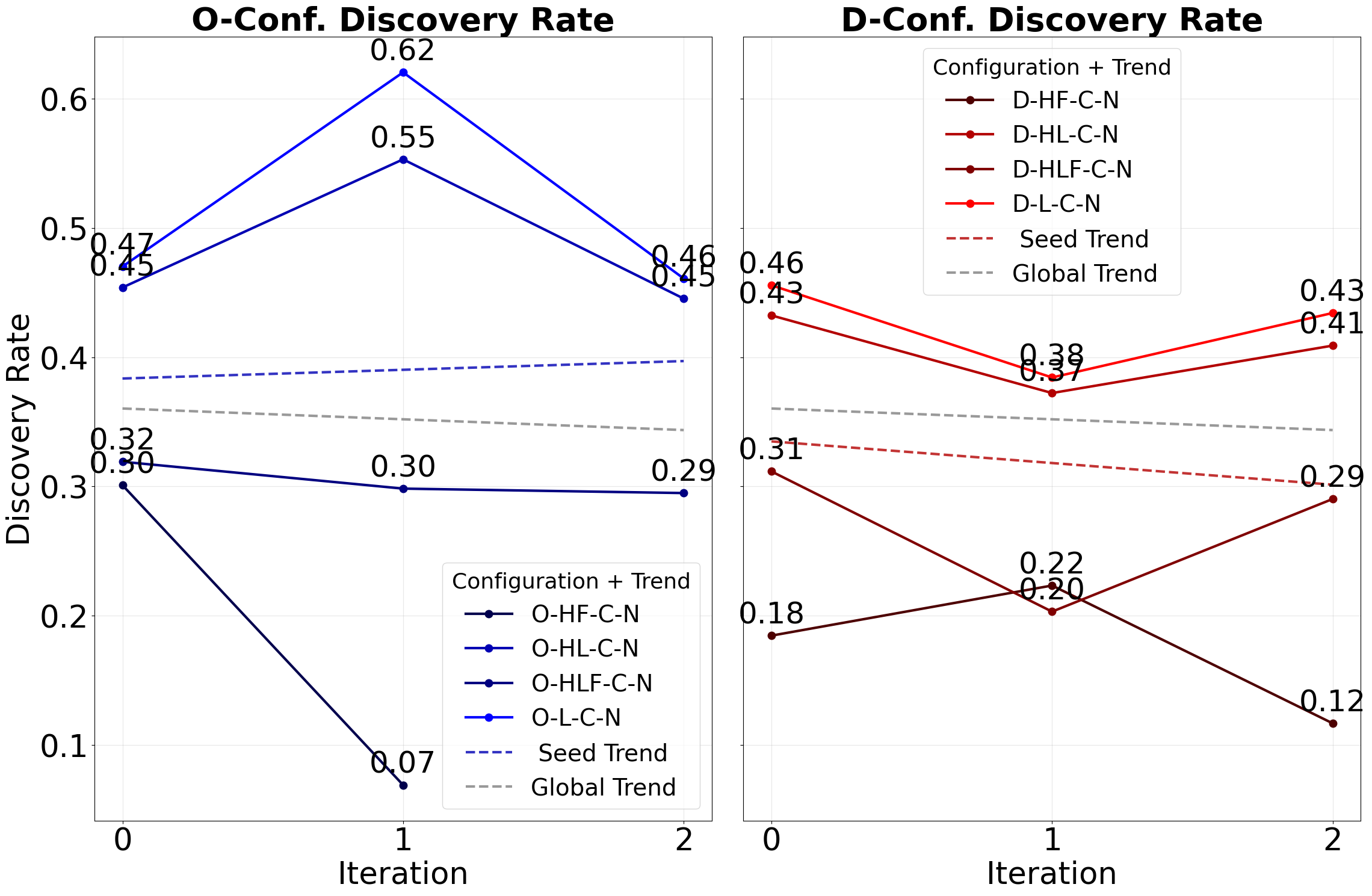}
        \caption{Discovery rate per iteration.}
        \label{fig:discovery_rate}
    \end{minipage}
    \hfill
    \begin{minipage}{0.49\columnwidth}
        \centering
        \includegraphics[width=\linewidth]{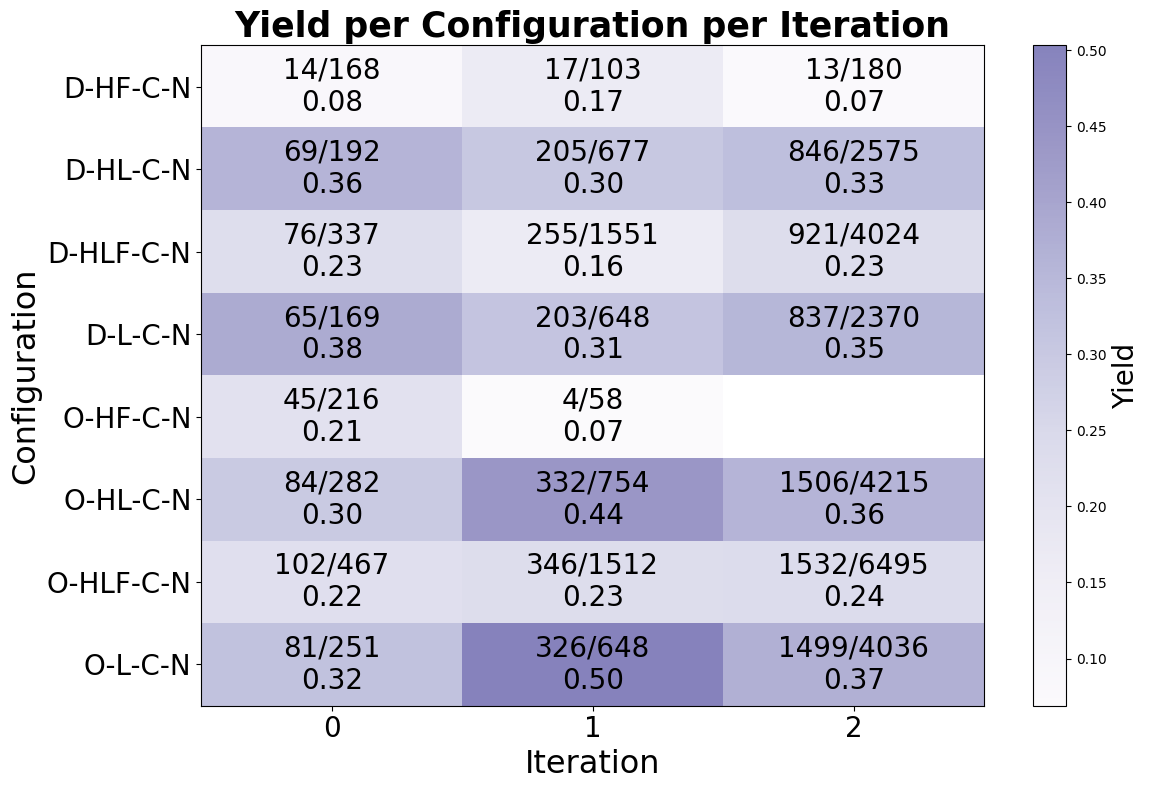}
        \caption{Yield per iteration.}
        \label{fig:yield}
    \end{minipage}

    \vspace{0.5em}

    \begin{minipage}{0.49\columnwidth}
        \centering
        \includegraphics[width=\linewidth]{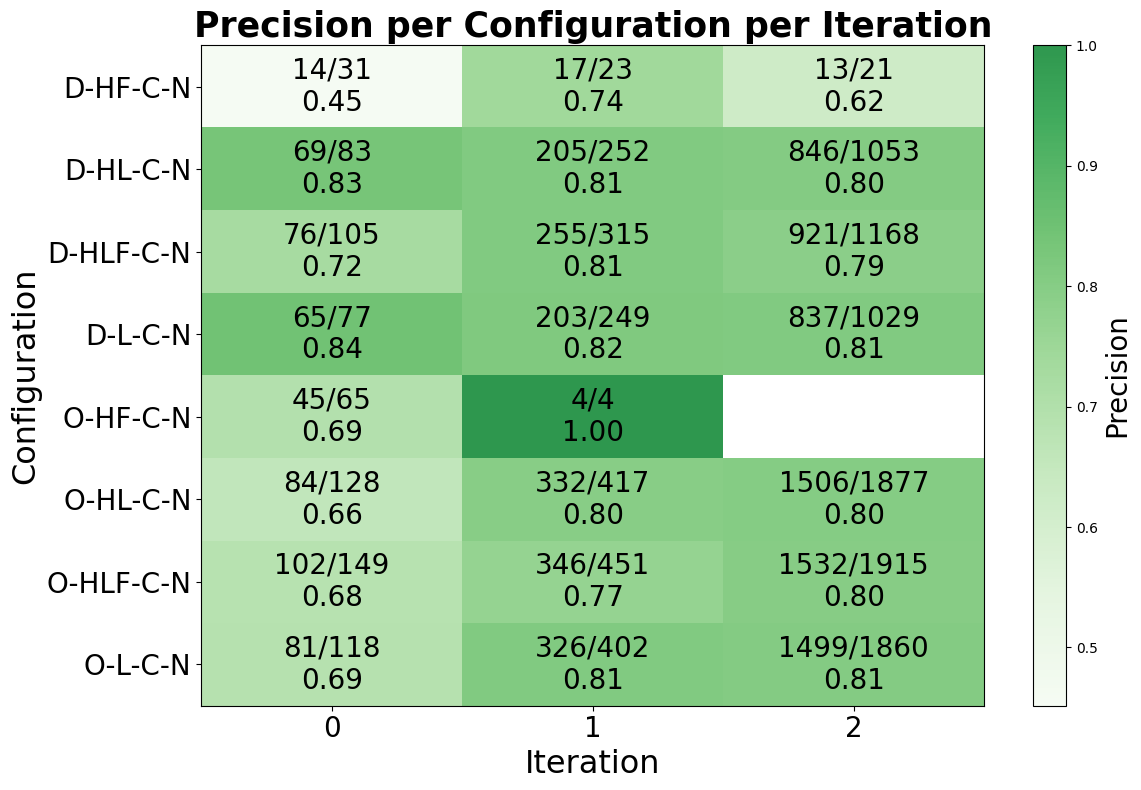}
        \caption{Precision per iteration.}
        \label{fig:precision}
    \end{minipage}
    \hfill
    \begin{minipage}{0.49\columnwidth}
        \centering
        \includegraphics[width=\linewidth]{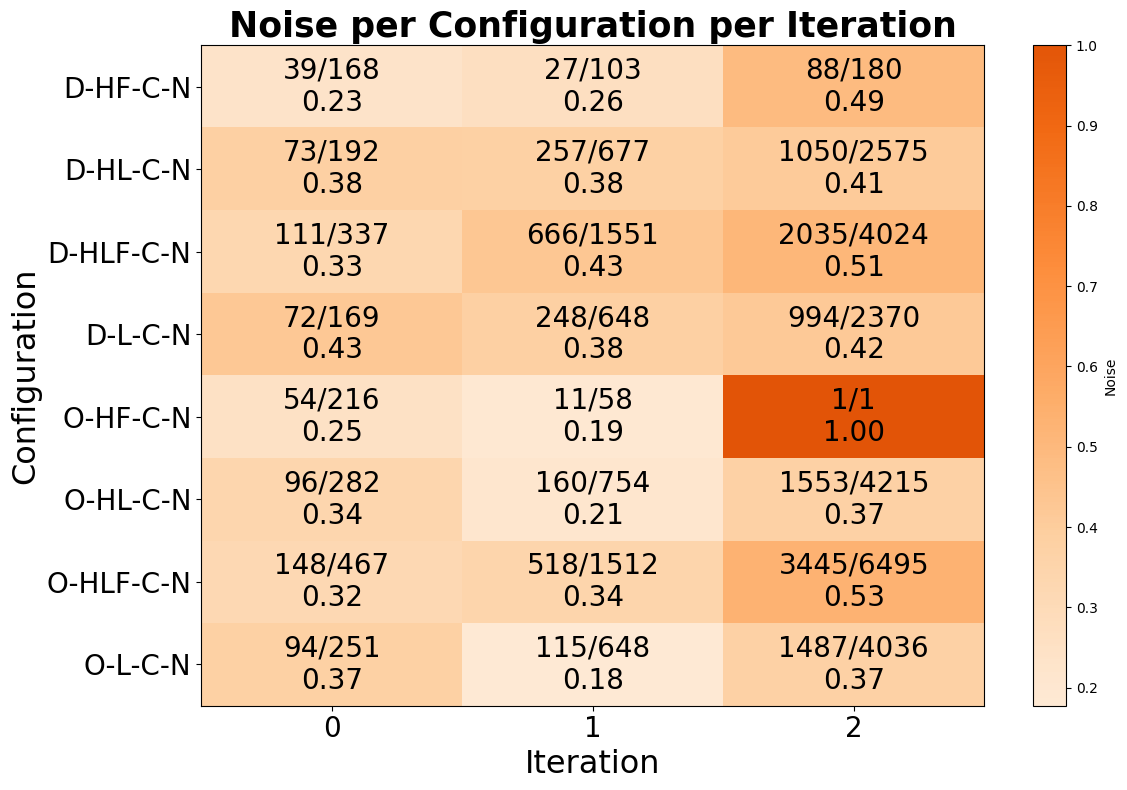}
        \caption{Noise per iteration.}
        \label{fig:noise}
    \end{minipage}
\end{figure}

Table~\ref{efficiency_results} summarises efficiency metrics across 
all configurations. Figures~\ref{fig:discovery_rate}--\ref{fig:noise} 
show per-iteration dynamics.

\textbf{Discovery rate.} Open Web configurations consistently 
outperform their Dark Web counterparts across all pointer types. The 
strongest configuration is \texttt{O-L-C-N}, achieving an average 
discovery rate of 0.52, compared to 0.42 for \texttt{D-L-C-N}. 
Similar patterns hold in link-aided settings: \texttt{O-HL-C-N} 
reaches 0.48 versus 0.40 for \texttt{D-HL-C-N}, and 
\texttt{O-HLF-C-N} reaches 0.30 versus 0.27 for \texttt{D-HLF-C-N}. 
Even in HF-only configurations the Open Web seed maintains a marginal 
advantage (0.18 vs.\ 0.17). Overall, Open Web configurations achieve 
discovery rates 3--24\% higher than equivalent Dark Web strategies, 
with open-web configurations displaying a slight upward trend across 
iterations while dark-web configurations show a mild decline 
(Figure~\ref{fig:discovery_rate}).

\textbf{Yield.} Yield varies primarily by pointer type, with 
link-based strategies consistently outperforming alternatives. 
Link-only configurations achieve the highest yields: \texttt{O-L-C-N} 
reaches 0.40 and \texttt{D-L-C-N} reaches 0.36, representing 
improvements of 18--25\% over HF-only settings (\texttt{O-HF-C-N}: 
0.34; \texttt{D-HF-C-N}: 0.19). Link-aided configurations (HL, HLF) 
exhibit similarly elevated yields. This advantage translates into 
output volume: \texttt{O-L-C-N} discovers 1,951 valuable communities, 
approximately 70\% more than \texttt{D-L-C-N} (1,145), and 
\texttt{O-HL-C-N} uncovers 1,967 compared to 1,160 for 
\texttt{D-HL-C-N}.

\textbf{Precision.} Precision remains consistently high across all 
strategies, generally 70--82\%, indicating that most newly discovered 
communities are genuinely valuable regardless of configuration. 
Link-only configurations achieve the highest overall precision 
(\texttt{D-L-C-N}: 0.82; \texttt{O-L-C-N}: 0.80) while operating at 
much larger scale than HF-only variants. \texttt{O-HF-C-N} also 
attains 0.82, but this reflects a substantially smaller search space 
(275 pointers processed); HF-only precision is thus somewhat 
artificially high due to early termination and shallow exploration 
depth. Link-based configurations maintain similarly high precision 
while scaling to much larger pointer and community volumes, reinforcing 
the central role of invite links in efficiently reaching valuable 
targets.

\textbf{Noise.} Noise represents the proportion of pointers that fail 
to resolve to Telegram communities (e.g., expired links, deleted 
accounts, malformed handles) and increases steadily across iterations 
for most configurations (Figure~\ref{fig:noise}). Noise frequently 
approaches or exceeds 40\% in link-based strategies (\texttt{D-L-C-N}: 
0.41; \texttt{O-L-C-N}: 0.34), with link-aided configurations 
accumulating even higher levels (\texttt{D-HLF-C-N}: 0.48; 
\texttt{O-HLF-C-N}: 0.49), reflecting the growing presence of 
outdated invite links during expansion. Dark Web seeds exhibit 
consistently higher noise overall, with all Dark Web configurations 
exceeding 0.34, whereas Open Web seeds maintain comparatively more 
stable resolution rates. As shown in Table~\ref{tab:regressions}, 
links reduce the odds of an invalid pointer by $\approx$90\% 
($\exp(-2.24)\approx0.11$), and each passing week increases noise by 
$\approx$50\% ($\exp(0.38)\approx1.46$), confirming pointer freshness 
as a key operational factor. Pointers shared in criminal contexts age 
better: each additional week makes them $\approx$14\% less likely to 
fail than equally old non-criminal pointers.

\begin{table*}[t]
\centering
\caption{Efficiency metrics across configurations (most effective in bold).}
\label{efficiency_results}
\resizebox{0.95\linewidth}{!}{
\begin{tabular}{l|rr|rr|rrrr}
\toprule
 & \multicolumn{2}{c|}{\textbf{Communities}} & 
   \multicolumn{2}{c|}{\textbf{Pointers}} & 
   \multicolumn{4}{c}{\textbf{Efficiency Metrics}} \\
\textbf{Configuration} & \textbf{Valuable} & \textbf{New Unique} & 
\textbf{Processed} & \textbf{Invalid} & 
\textbf{Avg.\ Disc.\ Rate} & \textbf{Yield} & 
\textbf{Precision} & \textbf{Noise} \\
\midrule
D-HF-C-N  & 84   & 118  & 451  & 154  & 0.17 & 0.19 & 0.71 & \textbf{0.34} \\
D-HL-C-N  & 1160 & 1431 & 3447 & 1382 & 0.40 & 0.34 & 0.81 & 0.40 \\
D-HLF-C-N & 1292 & 1631 & 5912 & 2812 & 0.27 & 0.22 & 0.79 & 0.48 \\
D-L-C-N   & 1145 & 1398 & 3187 & 1314 & \textbf{0.42} & \textbf{0.36} & \textbf{0.82} & 0.41 \\
\midrule
O-HF-C-N  & 94   & 114  & 275  & 66   & 0.18 & 0.34 & \textbf{0.82} & \textbf{0.24} \\
O-HL-C-N  & 1967 & 2467 & 5251 & 1809 & 0.48 & 0.37 & 0.80 & 0.34 \\
O-HLF-C-N & 2025 & 2560 & 8474 & 4111 & 0.30 & 0.24 & 0.79 & 0.49 \\
O-L-C-N   & 1951 & 2425 & 4935 & 1696 & \textbf{0.52} & \textbf{0.40} & 0.80 & 0.34 \\
\bottomrule
\end{tabular}
}
\end{table*}

\subsection{Accessibility of Discovered Communities (RQ2)}
\label{sec:results_accessibility}

\begin{table}[t]
\centering
\small
\caption{Community and access type per configuration.}
\label{accessibility_results}
\setlength{\tabcolsep}{4pt}
\renewcommand{\arraystretch}{0.95}
\begin{tabular}{lrrrrrr}
\toprule
 & \multicolumn{3}{c}{Comm.\ Type} & \multicolumn{3}{c}{Access Type} \\
Config. & Group & Chan. & Unk. & Pub. & Priv. & Vetted \\
\midrule
D-HF  & 20   & 98   & 3  & 117  & 3   & 1  \\
D-HL  & 822  & 604  & 48 & 1091 & 332 & 51 \\
D-HLF & 877  & 748  & 55 & 1268 & 359 & 53 \\
D-L   & 812  & 572  & 47 & 1055 & 325 & 51 \\
\midrule
O-HF  & 10   & 102  & 3  & 111  & 3   & 1  \\
O-HL  & 1349 & 1065 & 61 & 1902 & 502 & 71 \\
O-HLF & 1354 & 1160 & 61 & 1999 & 505 & 71 \\
O-L   & 1342 & 1029 & 61 & 1865 & 497 & 70 \\
\bottomrule
\end{tabular}
\end{table}

Public communities dominate all configurations (73--97\% of resolved 
communities), with Open Web seeds uncovering substantially more across 
all access types: \texttt{O-L-C-N} finds 1,865 public communities 
versus 1,055 for \texttt{D-L-C-N} (+77\%). Private communities 
comprise 20--22\% of link-aided Open Web results; vetted communities 
remain below 3.6\% in all configurations. These distributions reflect 
ecosystem accessibility as observable from our seeds and crawler 
visibility, and should not be generalized as a complete characterization 
of the Telegram cybercrime ecosystem.

\paragraph{Market segment distribution}

\begin{figure}[t]
    \centering
    \includegraphics[width=\columnwidth]{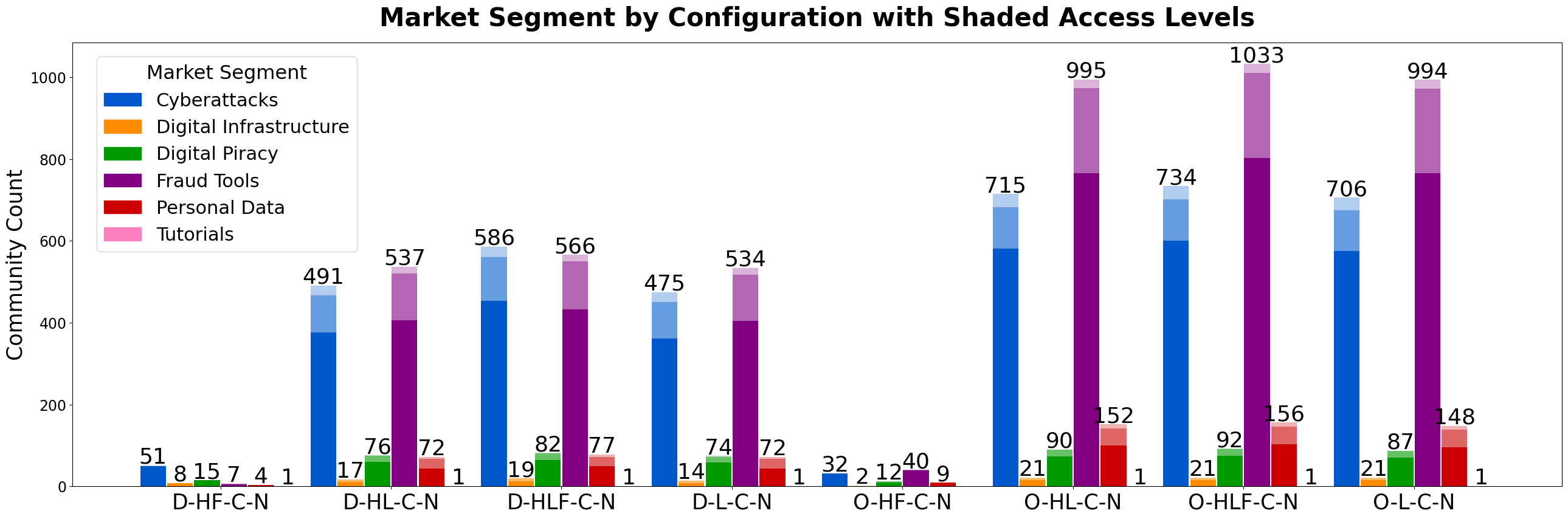}
    \caption{Market segment accessibility characteristics. Shading depicts public, private, and vetted communities from the bottom to the top.}
    \label{fig:accessdist}
\end{figure}

\begin{figure}[t]
    \centering
    \includegraphics[width=\columnwidth]{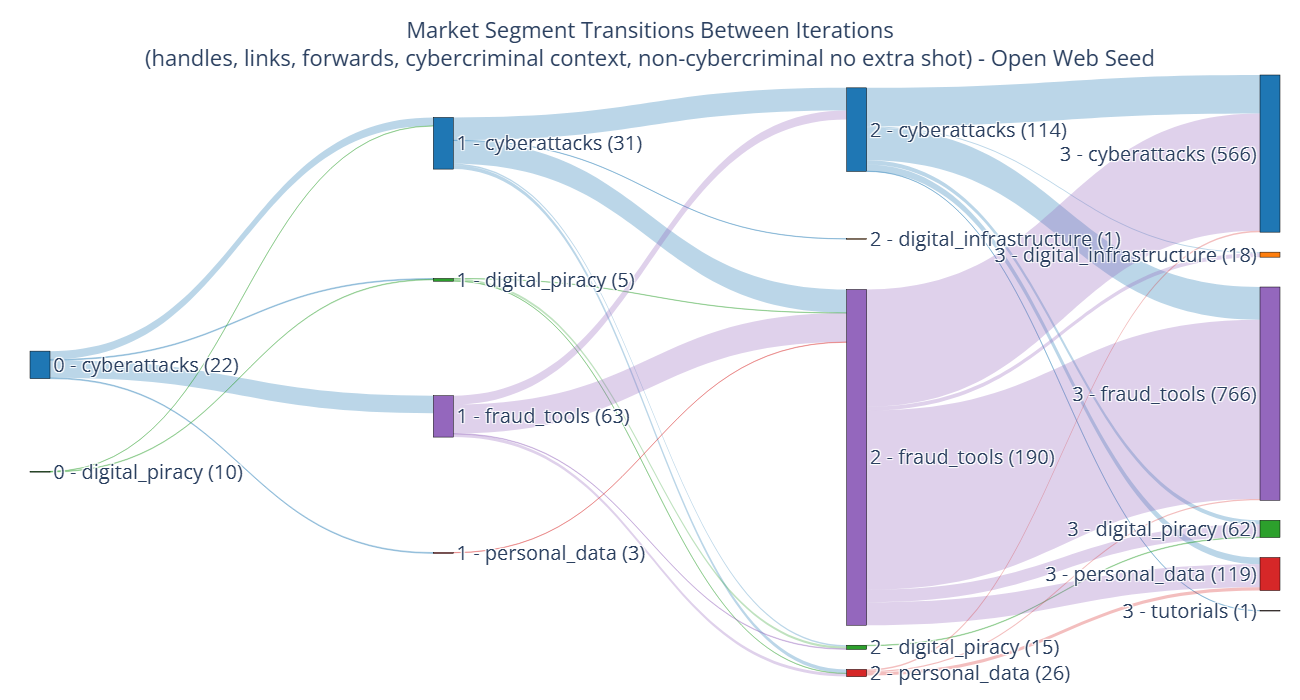}
    \caption{O-HLF-C-N access paths across segments.}
    \label{fig:sankey1}
\end{figure}

Figure~\ref{fig:accessdist} shows the distribution across market 
segments. \textit{Fraud Tools} and \textit{Cyberattacks} dominate 
discovery flows and function as structural hubs for cross-segment 
transitions. \textit{Digital Infrastructure}, \textit{Personal Data}, 
and \textit{Digital Piracy} appear less frequently; their weaker 
connectivity suggests greater specialization. Low counts for some 
segments (e.g., 2 Tutorial and 43 Digital Infrastructure communities) 
may partly reflect lower classifier recall for small classes in 
addition to true ecosystem structure. Figure~\ref{fig:sankey1} 
illustrates cross-segment paths under a link-enabled configuration: 
\textit{Cyberattacks} emerges as a central hub enabling transitions 
into less directly accessible segments, and link-based strategies 
produce broader cross-segment flows than handle/forward-only approaches.

\subsection{Rediscovery Extent (RQ3)}
\label{sec:results_rediscovery}

\begin{table}[t]
\centering
\small
\caption{Cumulative rediscovery rate across configurations.}
\label{rediscovery_results}
\setlength{\tabcolsep}{4pt}
\renewcommand{\arraystretch}{0.95}
\begin{tabular}{lrrr}
\toprule
Configuration & Unique Comm. & Redisc. & Cumm.\ Rate \\
\midrule
D-HF  & 118  & 3  & 0.02 \\
D-HL  & 1431 & 43 & 0.03 \\
D-HLF & 1631 & 49 & 0.03 \\
D-L   & 1398 & 33 & 0.02 \\
\midrule
O-HF  & 114  & 1  & 0.01 \\
O-HL  & 2467 & 8  & 0.00 \\
O-HLF & 2560 & 15 & 0.01 \\
O-L   & 2425 & 7  & 0.00 \\
\bottomrule
\end{tabular}
\end{table}

\begin{figure}[t]
    \centering
    \includegraphics[width=\columnwidth]{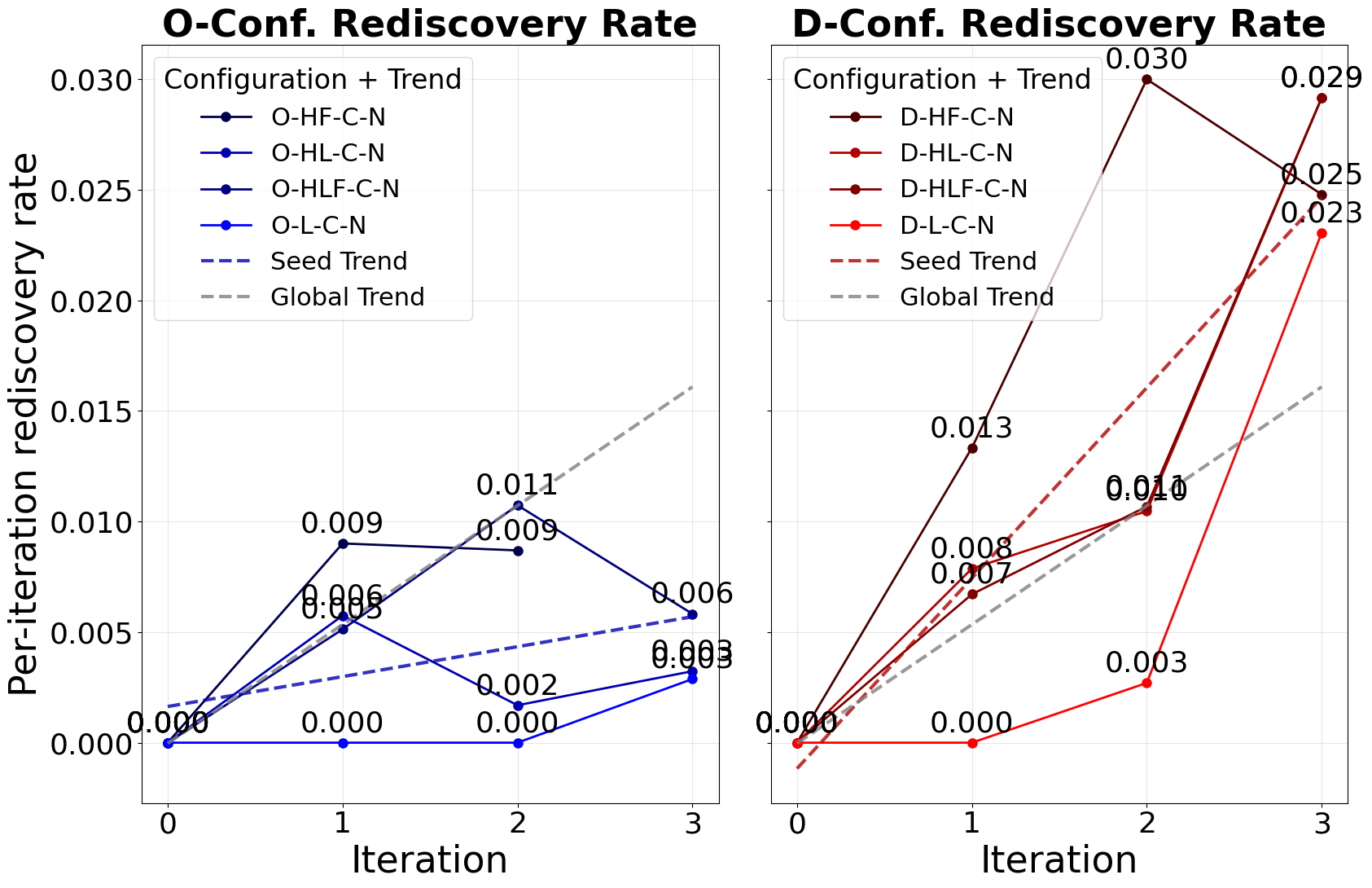}
    \caption{Per-visit redundancy across configurations.}
    \label{fig:rediscovery}
\end{figure}

Rediscovery remains low across all configurations 
(Table~\ref{rediscovery_results}), but dark-web seeds show higher 
redundancy (0.02--0.03) than open-web seeds ($\approx$0.00). 
Per-iteration redundancy (Figure~\ref{fig:rediscovery}) shows 
dark-web HL and HLF variants increasingly revisiting known communities 
by iteration 3, with the dark-web HLF redundancy reaching 0.029 — 
suggesting saturation may be beginning rather than absent under 
extended exploration. Open-web configurations maintain minimal 
redundancy throughout, indicating continued outward expansion.

\section{Discussion}
\label{sec:discussion}

Our results demonstrate that at-scale collection and discovery of 
criminal Telegram communities is possible, yet challenging to do efficiently without applying the appropriate techniques. 
Community discovery expands rapidly in absolute terms yet remains 
productive in uncovering cybercriminal communities only with certain configurations, suggesting that 
systematic measurements of this kind can generate meaningful insight 
into the Telegram underground and support efficient threat 
intelligence collection. The rapid expansion of discovered communities indicates that no single 
actor can realistically traverse the ecosystem exhaustively, pointing 
to a fragmented underground where actors are distributed across a 
large number of loosely connected channels and groups.

\textbf{Efficiency.} Discovery outcomes are primarily driven by 
link-based configurations, which consistently achieve the highest 
absolute discovery and sustain expansion across iterations. Noise 
remains substantial across all strategies (often exceeding 40\%), 
largely explained by pointer age. Seed provenance matters mainly when 
links are excluded; when links are incorporated, performance converges 
across open- and dark-web seeds, though open-web seeds generate higher 
absolute volumes. Dark-web seeds exhibit higher noise, reflecting a 
more volatile ecosystem where pointers decay or migrate more 
frequently. Pointers shared in criminal contexts appear more stable 
over time, suggesting that criminally embedded referrals may point to 
longer-standing communities.

\textbf{Accessibility.} Not all market segments are equally 
accessible. Fraud Tools and Cyberattacks dominate discovery flows and 
function as central transition hubs, consistent with their prominence 
in prior Telegram cybercrime 
studies~\cite{darkgram,article}. Segments such as Digital 
Infrastructure, Tutorials, and Personal Data appear less connected and 
more specialized. Compared to DarkGram~\cite{darkgram}, which 
characterized a fixed collection of 339 communities using seeded 
LLM-based classification, TeleHunt systematically evaluates how 
discovery configuration shapes what is found: we show that link-based 
strategies expose substantially broader segment diversity than 
handle/forward-only approaches, a dimension not examined in prior 
work. Our findings on accessibility are conditional on seed choice, 
pointer extraction rules, and crawler visibility, and should not be 
interpreted as a complete characterization of the ecosystem.

\textbf{Rediscovery.} Rediscovery rates remain consistently low, 
particularly for open-web seeds. Dark-web configurations show rising 
per-iteration redundancy by iteration 3, suggesting saturation may 
emerge under extended exploration rather than being categorically 
absent. The ecosystem appears continuously evolving, requiring 
sustained longitudinal monitoring to capture structural changes over 
time.

\subsection{Practical Implications}

The fraction of the Telegram cybercrime ecosystem reachable through our experiment appears highly discoverable and relatively loosely gated. For \textit{threat intelligence teams}, effective 
monitoring should prioritize link-centric collection and rapid 
ingestion to avoid decay-driven noise in later iterations. For 
\textit{law enforcement and CTI practitioners}, the predominance of 
public and lightly gated communities means substantial portions of the communities reachable through our exploration process remain directly accessible, with low rediscovery indicating 
continued opportunities to uncover previously unknown groups. For 
\textit{cybercrime researchers}, leveraging both open- and dark-web 
seeds is important to avoid structural bias toward dominant segments 
such as Fraud Tools and Cyberattacks, and the continued discovery of new communities within three iterations suggests that additional exploration depth may uncover further communities.

\subsection{Limitations}

Our results should not be interpreted as a full characterization of 
the Telegram cybercrime ecosystem, as visibility is bounded by initial 
seed choice. The two-week collection window captures only a snapshot 
of a fast-changing platform; longer collection periods would better 
capture enforcement cycles and community migration. Threshold values 
(70\%, 50\%, 30 days, 30 messages) were chosen on empirical 
distributional grounds; sensitivity analysis under alternative values 
is left for future work. Classification accuracy is affected by 
multilingual content, domain-specific slang, and class imbalance, 
increasing misclassification risk for niche segments — low discovery 
counts for some segments may partly reflect classifier recall rather 
than true ecosystem structure. Finally, Telegram's API rate limits, 
intermittent downtime, and shifting access models bound the 
scalability of at-scale collection.

\section{Conclusion}
\label{sec:conclusion}

TeleHunt enables configurable discovery, collection, and evaluation 
of cybercriminal Telegram communities. Link-based pointers drive 
discovery efficiency, outperforming handles and forwards, while 
context filtering and extra-shot expansion yield minimal additional 
gains (RQ1). Most discovered communities are public or lightly gated, 
with link-based exploration uncovering broader segment diversity; 
Fraud Tools and Cyberattacks appear most frequently in the explored discovery graph and exhibit stronger cross-segment connectivity than other segments (RQ2). Rediscovery 
rates are low, especially for open-web seeds, indicating limited evidence of saturation within the evaluated exploration depth (RQ3). The labeled 
dataset of 6{,}022 communities (3{,}471 cybercriminal), 
172{,}385{,}463 messages, 2{,}392{,}741 users, spanning 1{,}636 
Fraud Tools, 1{,}366 Cyberattacks, 244 Personal Data, 180 Digital 
Piracy, 43 Digital Infrastructure, and 2 Tutorial communities will 
be available for researchers that request access. Future work should pursue longitudinal 
analysis of the Telegram underground to track structural evolution, 
enforcement-driven migration, and the emergence of new market 
segments, and should examine sensitivity of discovery outcomes to 
threshold and seed choices.

\section*{Acknowledgements}

Part of this study is funded by the INTERSECT project,
Grant No. NWA.1162.18.301, funded by NWO and by the CATRIN project, Grant No. NWA.1215.18.003. 

\small
\bibliographystyle{splncs04}
\bibliography{references}

\appendix

\section{TeleHunt Framework Details}
\label{appendix:framework}

Key parameters about the snowballing configuration used in all 
experiments are displayed in Table~\ref{tab:parameters}. Rediscovery 
is prevented within runs by maintaining a log of previously visited 
communities. Full details of the handle reliability analysis and 
database schema are provided in our online repository \url{https://github.com/royricaldi/telehunt}.

\begin{table}[ht!]
\centering
\caption{TeleHunt snowballing configuration parameters.}
\label{tab:parameters}
\begin{tabular}{p{0.26\columnwidth} p{0.38\columnwidth} p{0.38\columnwidth}}
\toprule
\textbf{Parameter} & \textbf{Value} & \textbf{Purpose} \\
\midrule
Total iterations & 3 expansions + final classif. & Control growth and API load \\

Ad types & Handles, Invite Links, Forwards & Define traversal strategy \\

Ad age window & 30 days & Reduce stale/expired pointers \\

Dormant cutoff & <30 messages & Exclude inactive communities \\

Context window & varies (ad, replies) & Build context for filtering \\

Context filtering rule & $\geq$50\% cybercriminal in window & Reduce noisy pointers \\

Community threshold & $\geq$70\% cybercriminal messages & Define valuable communities \\

Both-context mode & Optional & Retain pointers failing default filter \\

Extra shot & Optional (1 iteration) & Expand via non-CC communities \\

\bottomrule
\end{tabular}
\end{table}

\section{Classification Models and Validation}
\label{appendix:classification}

Two RoBERTa-Large classifiers are used: (1) a \textbf{binary 
classifier} for Cybercriminal vs.\ Non-Cybercriminal and (2) a 
\textbf{segment classifier} assigning cybercriminal messages to six 
market segments or non-cybercriminal. Both were fine-tuned from 
\texttt{roberta-large-mnli} using 3-fold cross-validation.
Training data for the binary classifier consisted of 228 cybercriminal 
communities (285,895 messages) and a balanced non-cybercriminal sample 
(285,895 messages). For the segment classifier, 571,790 messages 
annotated with market segments were used, drawn from prior 
work~\cite{ricaldi_marjanov_hutchings_allodi_2025} and manually 
validated on a subset; inter-annotator agreement achieved F1\,=\,0.871 
and Cohen's $\kappa$\,=\,0.843.

\begin{table}[ht!]
\centering
\small
\caption{Binary classifier performance (avg.\ across epochs and folds).}
\label{tab:binary-classifier}
\setlength{\tabcolsep}{4pt}
\renewcommand{\arraystretch}{0.95}
\begin{tabular}{lcccc}
\toprule
\textbf{Class} & \textbf{Prec.} & \textbf{Rec.} & 
\textbf{F1} & \textbf{Support} \\
\midrule
Cybercriminal     & 0.9923 & 0.9881 & 0.9902 & 95,298 \\
Non-Cybercriminal & 0.9881 & 0.9923 & 0.9902 & 95,297 \\
\midrule
Macro / Weighted  & 0.9902 & 0.9902 & 0.9902 & 190,595 \\
\bottomrule
\multicolumn{5}{l}{\footnotesize Overall accuracy: 0.9902.}
\end{tabular}
\end{table}

\begin{table}[ht!]
\centering
\small
\caption{Segment classifier performance (avg.\ across epochs and folds).}
\label{tab:segment-classifier}
\setlength{\tabcolsep}{4pt}
\renewcommand{\arraystretch}{0.95}
\begin{tabular}{lcccc}
\toprule
\textbf{Class} & \textbf{Prec.} & \textbf{Rec.} & 
\textbf{F1} & \textbf{Support} \\
\midrule
Cyberattacks           & 0.9512 & 0.9649 & 0.9578 & 30,373 \\
Digital Infrastructure & 0.9058 & 0.8820 & 0.8931 &  3,764 \\
Digital Piracy         & 0.9395 & 0.9337 & 0.9364 & 11,211 \\
Fraud Tools            & 0.9640 & 0.9516 & 0.9576 & 38,470 \\
Non-Cybercriminal      & 0.9822 & 0.9887 & 0.9854 & 14,605 \\
Personal Data          & 0.9321 & 0.9402 & 0.9360 & 11,255 \\
Tutorials              & 0.8458 & 0.8320 & 0.8348 &    225 \\
\midrule
Macro Avg              & 0.9315 & 0.9276 & 0.9287 & 109,903 \\
Weighted Avg           & 0.9549 & 0.9546 & 0.9545 & 109,903 \\
\bottomrule
\multicolumn{5}{l}{\footnotesize Overall accuracy: 0.9546. 
Weighted F1: 0.9545. Macro F1: 0.9287.}
\end{tabular}
\end{table}

\subsubsection{Human Validation}
\label{app:humanvalidation}

To assess the reliability of automated community classification, we 
manually validated a stratified sample of 100 communities per dataset 
(dark web and open web), preserving the predicted market segment 
distribution. For each community, up to 30 messages were randomly 
selected and inspected to determine the dominant market segment. 
Human annotations were compared with automated labels using exact 
agreement and Cohen's $\kappa$. Results show substantial agreement: for the dark web dataset, exact 
agreement reached 74\% ($\kappa=0.69$); for the open web dataset, 
71\% ($\kappa=0.65$). Macro-averaged F1-scores were 0.73 (dark web) 
and 0.70 (open web). Disagreements occurred primarily between closely 
related segments, particularly Cyberattacks and Fraud Tools.

\end{document}